# Variable Earth's Rotation Speed in the 14th to 16th Centuries: New ΔT Constraints from Chinese Eclipse Records

Hisashi Hayakawa (1 – 4), Mitsuru Sôma (5), Naiqi Li (2, 6)

(1) Institute for Space-Earth Environmental Research, Nagoya University, Nagoya, 4648601, Japan

(2) Institute for Advanced Research, Nagoya University, Nagoya, 4648601, Japan

(3) Space Physics and Operations Division, RAL Space, Science and Technology Facilities Council, Rutherford Appleton Laboratory, Harwell Oxford, Didcot, Oxfordshire, OX11 0QX, UK

(4) Nishina Centre, Riken, Wako, 3510198, Japan

(5) National Astronomical Observatory of Japan, Mitaka, 1818588, Japan

(6) Graduate School of Humanities, Nagoya University, Nagoya, 4648601, Japan

**Abstract**

Total solar eclipses are not only astronomical spectacles but also great astrophysical laboratories. Their historical records are particularly helpful for assessing the past variability of the Earth's rotation speed. Chinese records played a key role for such analyses. However, Chinese eclipse records from the Míng period have not been used for ΔT reconstructions, partially because most of the contemporaneous eclipse reports are found not in official histories but in local treatises. This study examines eclipse records in the (quasi-)contemporaneous local treatises, concentrating on what explicitly mentioned eclipse totality on the day of a total solar eclipse and what were compiled during the Míng Dynasty. On their basis, our study revised the ΔT constraint in 1361 to $-408$ s ≤ ΔT ≤ 601 s and set new ΔT constraints of 277 s ≤ ΔT ≤ 890 s in 1514, $-328$ s ≤ ΔT ≤ 332 s in 1542, and $-1762$ s ≤ ΔT ≤ 1091 s in 1575, respectively. We also revised most of the existing ΔT constraints in the 14th to 16th centuries, using the ephemeris data of the NASA JPL DE 441. Overall, our ΔT constraints generally tighten the ΔT variations more than what M+21 fit for their ΔT spline curve, requiring downward modification and upward modifications for the ΔT reconstructions around 1361 and 1542, respectively. Our results suggest that the ΔT decrease between 1514 and 1567 was slightly steeper than previously considered.

## 1. Introduction

Total solar eclipses have served human beings not only as one of the greatest astronomical spectacles, but also as one of the unique astrophysical laboratories (Orchiston et al., 2015; Littmann and Espenak, 2017; Pasachoff, 2017). Their long-term records have been used for multiple scientific measurements such as those for the solar coronal structures (Loucif and Koutchmy, 1989; Riley et el., 2015; Hayakawa et al., 2021, 2025) and solar diameters (Fiala et al., 1994; Rozelot and Damiani, 2012). These records have also served as vital references to assessing and reconstructing the long-term variability of Earth's rotation speed before the 1620s (Stephenson, 1997, hereafter S97; Sôma and Tanikawa, 2015; Orchiston et al., 2015; Stephenson et al., 2016, hereafter SMH16; Morrison et al., 2021, hereafter M+21; Hayakawa et al., 2022, 2024, 2025; Martínez Usó and Marco Castillo, 2023).

For such assessments and reconstructions, we owe a lot to Chinese historical records, as they have accommodated numerous reports of deep solar eclipses in the past and formed one of the backbones for the ΔT reconstructions (S97; Stephenson *et al.*, 2018; Hayakawa et al., 2025). Among them, eclipse reports in official histories (正史) down to the Yuán (元) Dynasty (1271 – 1368) have been of particular importance and have been intensively studied by Stephenson et al. (2018) in the context of the ΔT reconstructions. In contrast, Chinese eclipse records from the Míng (明) Dynasty (1368 – 1644) have not been used for ΔT reconstructions, partially because most of the contemporaneous eclipse reports are found not in official histories but in local treatises (地方志). S97 (pp. 258-262) documented the absence of eclipse reports in the treatises of *Míngshǐ* (明史), a lack of any allusion to large eclipses in the imperial annals of *Míngshǐ*, and difficulty in locating the observational sites, mentioning four total solar eclipses during the Míng period with multiple reports from local treatises. While Han and Qiao (2003) once studied the 1542 eclipse records on their basis, such local records have been mostly forgotten in the recent ΔT reconstructions (SMH16; M+21).

However, the Chinese eclipse records from the Míng periods may not have been as useless as previously considered. First, *Míngshǐ*, the official history of the Míng Dynasty, actually hosts a report of a total solar eclipse in 1575 that S97 missed. Moreover, the local treatises were designed to document local details such as geography, history, and events for the specific location to which their titles are dedicated. S97 himself used *Sōngjiāngfŭzhì* (松江府志) twice as source reports with a definite observational site (Sōngjiāngfŭ near Shànghǎi). In fact, not only S97 but also SMH16 and M+21 used *Sōngjiāngfŭzhì* to set a critical ΔT limit in 1361, although their interpretations are not always the same (see Section 3). More than a few of the Chinese local treatises are dedicated to the units of fŭ (府) and their subdivisions (*e.g.*, Hucker, 1958, pp. 44-45) that S97 has also considered specific enough for the geographical identification of an observational site. We also need to be mindful whether their source reports were from eyewitness accounts, as the vast majority of what Han and Qiao (2003) studied for the 1542 eclipse was coming from Qīng (清) period, at least more than a century after the eclipse in question, and hence can have easily been affected by the transmission process of hearsays over generations.

In this study, we explore such Chinese eclipse reports in the 14th to 16th centuries and derive their ΔT constraints for each total solar eclipse. After documenting source records (Section 2) and our methodology (Section 3), we develop case studies on the individual eclipses in 1361 (Section 4), 1514 (Section 5), 1542 (Section 6), and 1575 (Section 7), where Chinese local treatises explicitly mention the local total obscurations. We then revise most of the ΔT constraints that M+21 used for their ΔT reconstructions in the 14th to 16th centuries (Section 8). We contextualise our results with the existing ΔT reconstructions around the 14th to 16th centuries. Adding such datasets, this study offers additional ΔT constraints to tighten and improve the ΔT variations in the 14th to 16th centuries and further discussions on the long-term variability of the sea level, the global ice amounts, and the core-mantle interactions (Lambeck et al., 2014; Mitrovica et al., 2015; Zhu et al., 2021; Rekier et al., 2022; Suttie et al., 2025).

**2. Source Documentation: Local Treatises**

As S97 (pp. 260-262) discussed, official histories of Míng Dynasty do not offer many eclipse records. In contrast to S97, we found one such report in *Míngshĭ* (R7-5), although this report is coming from an imperial chronicle (本紀) and does not guarantee the observational site as the imperial capital.

In fact, many of the Chinese astronomical records from this period came from local treatises (S97). Local treatises (地方志), which are occasionally translated as "local gazetteers" or "provincial histories", are described as "chronicles of the history, present conditions, and noted people of local areas, arranged by topics" that are "compiled for administrative units, from the canton and county up to the provincial level and finally the country as a whole" (Brooks, 1988, p. 49). In addition, local treatises do not only concern administrative units; there are also treatises devoted to natural features (*e.g.*, treatises for mountain or lake) and to specific sites (*e.g.*, treatises for historical remains or bridges) (Ba, 2015).

In contrast with official histories, local treatises were written by local gentry, such as local literati, retired officials, and local magistrates (Brooks, 1988, p. 55). Staying out of imperial observatories, government astronomers were not likely involved, contrary to some speculations. Subsequently, these records were not as reliable as official histories (S97, p. 260), as government astronomers did not offer their observations to these records. Some of these appear to have been copied from the previous documents. However, the dates were not always copied accurately. We can easily find such evidence in the discrepancy between the dates and locations of some of their eclipse reports and the modern eclipse calculations (S97, p. 260). For example, R4-1[1] was most likely derived from R4-2, whereas it correctly conveyed the date without alteration. More problematically, we have seen duplicate reports with data problems. For example, *Xiùshuǐxiànzhì* (秀水縣志, v. 10, f. 3a; see R7-2) involves two entries with exactly the same details for the eclipses in 1542 and 1575. Beijing Observatory (1988) omitted the 1542 entries and included only the 1575 entries for a good reason. We agree with this interpretation, as this local treatise was compiled in Wànlì 24th year (≈ 1596) and the author has a higher probability of having an experience or a direct hearsay of the 1575 eclipse rather than the 1542 eclipse.

Owing to their motivations, these local treatises generally describe the local events that occurred locally in the administrative units to which their contents were dedicated. S97 associated *Sōngjiāngfǔzhì* (松江府志) twice with *Sōngjiāngfǔ* (Sōngjiāng Province) as a definite observational site. In the Míng administration, the provincial administration hierarchy descends from provinces (府, fǔ), subprefectures (州, zhōu), and counties (縣, xiàn), as reviewed in Hucker (1958, pp. 44-45). Therefore, any local treatises that are dedicated to provinces (fǔ), subprefectures (zhōu), and counties (xiàn) define their observational sites as geographically as specific as or more specific than *Sōngjiāngfǔzhì* from which S97 (p. 261) derived a "definite site" of the eclipse observation.

[1] Here, we put a record ID as R Section Number – Record Number. For example, R4-1 and R7-2 mean the first record of Section 4 and the second record of Section 7.

## 3. Materials and Methods

In order to satisfy the said purpose, we first picked up the dates for total solar eclipses that were viewed in China during the reigns of the Yuán and Míng Dynasties, using the *NASA Five Millennium Catalog of Solar Eclipses* (Espenak and Meeus, 2009). We then compared their dates to the union catalog of Chinese astronomical records (中國古代天象記錄總集; Beijing Observatory, 1988) to list the total solar eclipses for which Chinese reports indicated the visibility of total obscurations with the technical term of jì (既). This terminology has been used since *Chūnqiū* (春秋), at least (S97, pp. 223-224). In addition to the marginal eclipse case on 1361 May 5, we identified three such total solar eclipses on 1514 Aug 20, 1542 Aug 11, and 1575 May 10, where the local visibility of total solar eclipses was confirmed both from the *NASA Five Millennium Catalog of Solar Eclipses* (Espenak and Meeus, 2009) and the union catalog of Chinese astronomical records (Beijing Observatory, 1988, hereafter BO88).

For each case, we consulted the original records on which BO88 relied. We specifically examined eclipse records with two criteria. The first criterion is for these records to have used the technical term of jì. This technical term indicates total obscuration in the astronomical context.

The second criterion was that their source records were compiled during the reign of the Míng Dynasty. These records are generally coming from local treatises (地方志) that were written to describe local details for specific places including omens (祥異) such as celestial phenomena. Their purposes allow us to reasonably consider these originated in the target regions. Nevertheless, it is better to select records that are sufficiently close to those of the original eyewitnesses to make their descriptions as reliable as possible. We omitted local treatises that were compiled during the Qīng period in contrast with Han and Qiao (2003), as the Qīng rule over the China mainland started from 1644, some 139 years after the latest eclipse event that we discuss. The human lifespan is too short to let eyewitnesses of these eclipses survive until the Qīng period. These local treatises relied the source of these total solar eclipses at best on their hearsay and do not offer the best reliability.

We then located the observational sites of our reports on the basis of their regions to which the local treatises were dedicated in their titles. We consulted the *Comparison Table of Chinese Ancient Place Names* (中国古今地名對照表; Xue, 2009, hereafter X09) to associate the historical place names with their modern counterparts. We then extracted the geographical coordinates according to satellite imageries from Google Earth Pro. When X09 did not allow us to associate historical place names with their modern counterparts, we searched for places with the same name in the same province on Google Earth Pro.

Upon identification, we computed the local eclipse visibility using the ephemeris data of NASA JPL DE 441 (Park et al., 2021), running the calculation code of Sôma and Tanikawa (2015), and changing the parameter of the Earth's rotation speed. We first used the values of M+21's ΔT spline curve to check the local eclipse visibility. We then changed the ΔT parameter to search for the boundary conditions that allowed each observational site to see a local total obscuration. We then searched for the lower and upper ΔT limits to locate all such sites in the totality path for each target solar eclipse. The results are summarised in Table 1. Our results are compared with M+21's ΔT spline curve and contemporaneous ΔT constraints.

**4. Total Solar Eclipse of 1361 May 5**

This eclipse is known to S97 (pp. 259-260) through *Sōngjiāngfŭzhì* (松江府志), while BO88 listed another record (*Náncūn Chuògēnglù*; 南村輟耕錄) that indicated a totality. S97 and SMH16 regarded the former as a report for deep partial solar eclipse, whereas Morrison et al. (2020) and M+21 changed their interpretation to that of a total solar eclipse following their private communication with David Dunham. SMH16 located their spline curve at ΔT < 500 s in 1361, while M+21 located their ΔT margin of 1361 in 500 s < ΔT < 1760 s. This discrepancy requires revisiting the source report(s). The results are as follows.

R4-1: *Sōngjiāngfŭzhì* (松江府志, v. 47, f. 19a)
Our Transcription: 辛丑四月朔日，日將沒，忽無光，作蕉葉樣。天黑如夜，星斗燦然。食頃，天再明，又少時，乃沒。
Our Translation: On Xīnchŏu [year], the fourth month, the first day; the Sun was about to set, suddenly lost light, and took on a shape of plantain leaf. The sky darkened as at night. Star(s) shone brightly. After a short period, the sky brightened again. After another short period, the Sun finally sets.

R4-2: *Náncūn Chuògēnglù* (南村輟耕錄, v. 19, f. 17b)

Our Transcription: 至正辛丑四月朔日，日未沒三四竿許，忽然無光，漸漸作蕉葉樣。天且昏黑如夜，星斗粲然。飯頃，方復舊，天再開，星斗亦隱。又少時，乃沒。

Our Translation: In Zhìzhèng Xīnchǒu [year], the fourth month, the first day; the Sun had not yet set and was still about three to four gān[2] above the horizon.. The Sun suddenly lost light and gradually assumed the shape of a plantain leaf. The sky darkened as at night. Star(s) shone brightly. After a short period, the Sun returned to its original form and the sky brightened again. Star(s) were also hidden. After another short period, the Sun finally set.

As can be seen from their translations, they share key details about this eclipse. Both located the eclipse maximum immediately before sunset, compared the shape with a plantain leaf, and described a short star appearance(s). R4-1 is obviously dedicated to Sōngjiāngfǔ that has a preface dated in Chóngzhēn 4th year (≈ 1631). R4-2 is an essay that Táo Zōngyí (陶宗儀) wrote through his life in Sōngnán (松南) of Sōngjiāngfǔ and accumulated in pots in parallel with his agricultural works (*Náncūn Chuògēnglù*, v. P. f. 1a). Sōngnán is located in Sìjīng (泗涇), according to *Míngshīzōng* (明詩綜, v. 12, f. 1b). *Náncūn Chuògēnglù* (v. P, f. 2b) has a preface dated in 1366. Their similarity and chronological offsets make it difficult to consider them independent of each other, and allow us to reasonably conclude that R4-2 is a source of R4-1 in this specific case. From the date of the preface, it is reasonable to regard Táo Zōngyí as an eyewitness to this eclipse.

[2] A traditional unit indicating the height of a pole.

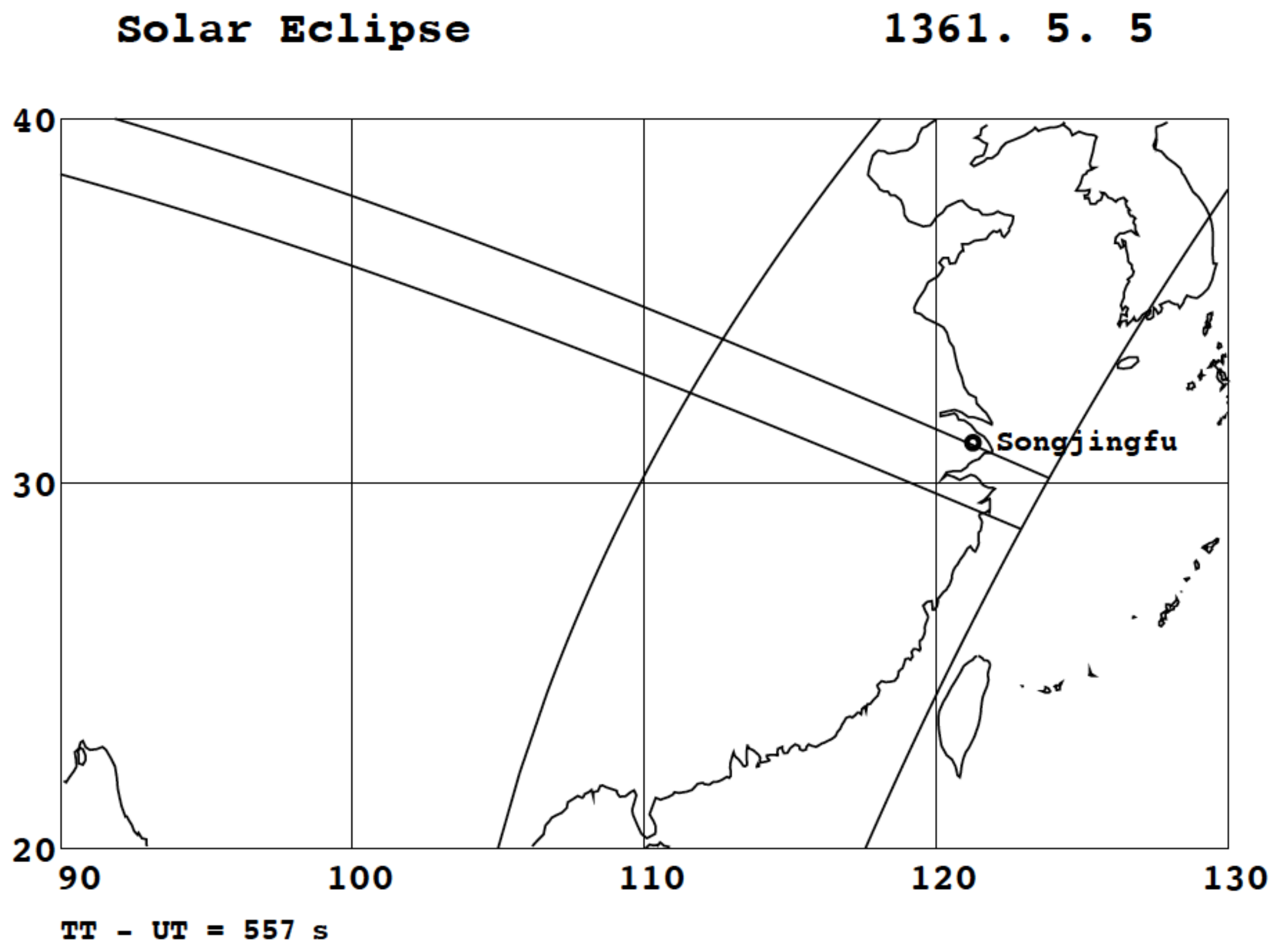

Figure 1: Our calculation of the totality path of the 1361 total solar eclipse in comparison with the location of Sōngjiāngfǔ, where we used M+21's ΔT value (ΔT = 557 s) in 1361.

We located Sìjīng at N31°07′, E121°16′ in the modern geographical coordinates. Using M+21's ΔT spline curve (ΔT = 557 s in 1361), our calculation located Sijīng outside the totality path (Figure 1) and allowed this eclipse to only reach a maximal magnitude of 0.998. In order to locate Sìjīng in the totality path, we need to locate the ΔT in 1361 in the range of 602 s ≤ ΔT ≤ 1864 s. His description compared the shape of the eclipsed Sun with a plantain leaf and reminds us of a great partial solar eclipse, as S97 (pp. 259-260) also pointed out. While we need independent records to be sure, this description favours locating Sìjīng outside the totality path and setting the ΔT constraint in 1361 as ΔT ≤ 601 s, in accordance with SMH16 and in contrast with M+21.

Additionally, both of these records located a brightening before the local sunset. The Sun anyways set eclipsed, as long as we expect a ΔT value in 1361 in accord with contemporaneous ΔT constraints (−3547 s ≤ ΔT ≤ 2771 s). These descriptions allow us to locate the eclipse maximum before the local sunset. In this case, we need to set a lower ΔT limit of −408 s ≤ ΔT. The eclipse maximum was somewhere near the local sunset and at least after the local noon. In order to locate the eclipse maximum after the local noon, we need to set an upper an upper ΔT limit of ΔT ≤ 18532 s. In combination, we need to set a ΔT constraint of −408 s ≤ ΔT ≤ 601 s or 1865 s ≤ ΔT ≤ 18532 s. In reality, the ΔT value should have been much closer to 400 – 600 s, as we need a great eclipse magnitude to satisfy the said descriptions.

**5. Total Solar Eclipse of 1514 August 20**

For this eclipse, we found two reports that explicitly mentioned totality and were compiled during the Míng period. BO88 also listed *Qiānshū* (鉛書) that satisfies our criteria, while we could not locate this source in the TNL or Kanseki Database. We have put this report aside, as we could not verify the details. These two records are read as follows:

R5-1: *Jiǔjiāngfǔzhì* (九江府志, v. 1, f. 18a)
Our Transcription: 正德九年八月朔，日食之既，晝晦，星見，鷄犬驚鳴。
Our Translation: Zhèngdé 9th year, 8th month, the first day, the Sun was completely eclipsed. The daytime darkened so that star(s) were visible. The chickens and dogs were surprised and cried out.

R5-2: *Dōngxiāngxiànzhì* (東□縣志, v. 2, f. 42b)
Our Transcription: 正德九年甲戌八月朔，是日天高氣爽，四無雲翳。午時忽日食既，星見晦瞑，咫尺不辨，鷄犬驚宿，人民駭懼。歷一時，復明。
Our Translation: Zhèngdé 9th year Jiǎxū, 8th month, the first day; on this date, the sky was high and the air was refreshing. No cloud covers was observed in any directions. At wǔ, the Sun suddenly became totally eclipsed. Star(s) were visible. It was so dark that people could not distinguish objects even a zhǐ or a chǐ away. The chickens and dogs were surprised and roosted in their nest. People became terrified. After a double hour, the light returned.

These records commonly describe star appearances and animal reactions. R5-2 is more detailed about the human reactions and the emergence. R5-1 and R5-2 have prefaces dated in Jiājìng 6th year (≈ 1527) and Jiājìng 3rd year (≈ 1524), respectively. They were dated only slightly more than a decade after the total solar eclipse in 1514. It is reasonable to expect that either their authors or their direct informants were eyewitnesses of this eclipse.

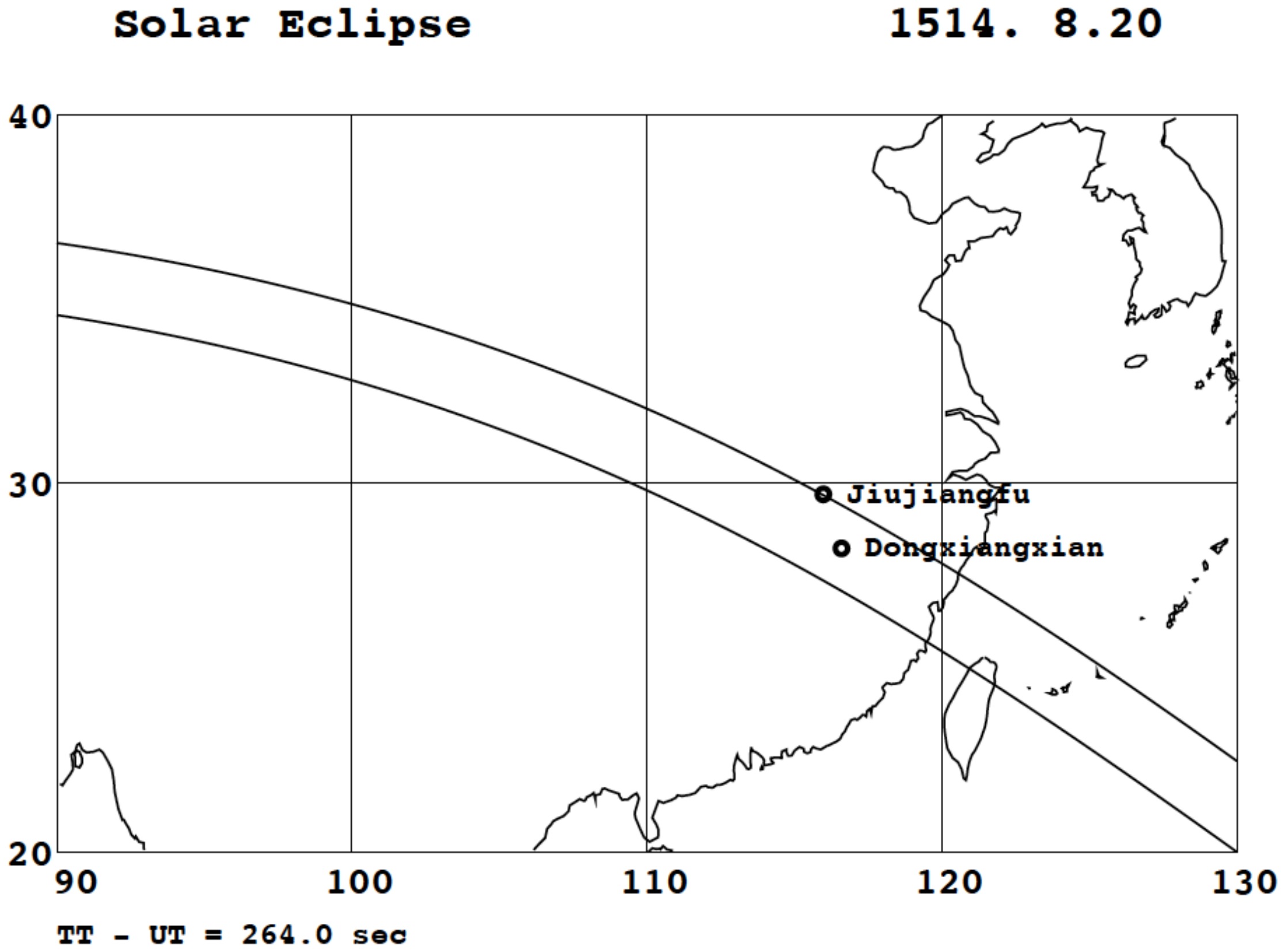

Figure 2: Our calculation of the totality path of the 1514 total solar eclipse in comparison with the location of Jiŭjiāngfŭ and Dōngxiāngxiàn, where we used M+21's ΔT value (ΔT = 264 s) in 1514.

Combining X09 and Google Earth Pro, we located Jiŭjiāngfŭ at N29°42′, E115°59′ and Dōngxiāngxiàn at N28°14′, E116°36′ in the modern geographical coordinates, respectively. Using M+21's ΔT spline curve (ΔT = 264 s in 1514), our calculation located Dōngxiāngxiàn in the totality path but Jiŭjiāngfŭ outside the totality path (Figure 2). In this case, this eclipse reached a maximal magnitude of 0.999. In order to locate both Jiŭjiāngfŭ and Dōngxiāngxiàn in the totality path, we need to locate the ΔT in 1514 in the range of 277 s ≤ ΔT ≤ 890 s. Even if we follow BO88, add Qiānshū to our consideration, and locate the site at Qiānshānxiàn (鉛山縣; N28°18′, E117°43′), Qiānshānxiàn was located in the totality path in the ΔT margin of −46 s ≤ ΔT ≤ 1195 s in 1514 and did not conflict with our results. This result allows us to add a new ΔT constraint in 1514 and requires us to revise the ΔT spline curve slightly upward from what M+21 published (ΔT = 264 s in 1514).

## 6. Total Solar Eclipse of 1542 August 11

For this eclipse, we found one report that explicitly mentioned the totality, which was compiled during the Míng period. BO88 also listed *Jiǔjiāngfǔzhì* (儀封縣志) that satisfies our criteria, while we could not locate this source in the TNL or Kanseki Database. We have put this report aside, as we could not verify the details. This record is as follows:

R6-1: *Shūchéngxiànzhì* (舒城縣志, v. 29, f. 12a)
Our Transcription: 嘉靖…二十一年七月朔，日食既，大星晝見。
Our Translation: Jiājìng ... 21th year, 7th month, 1st day, Sun was totally eclipsed. Large star(s) were observed during the daytime.

This record describes star appearances. R6-1 has a preface dated in Wànlì 8th year (≈ 1580). This record was published more than three decades after this eclipse. This account is probably based on hearsay.

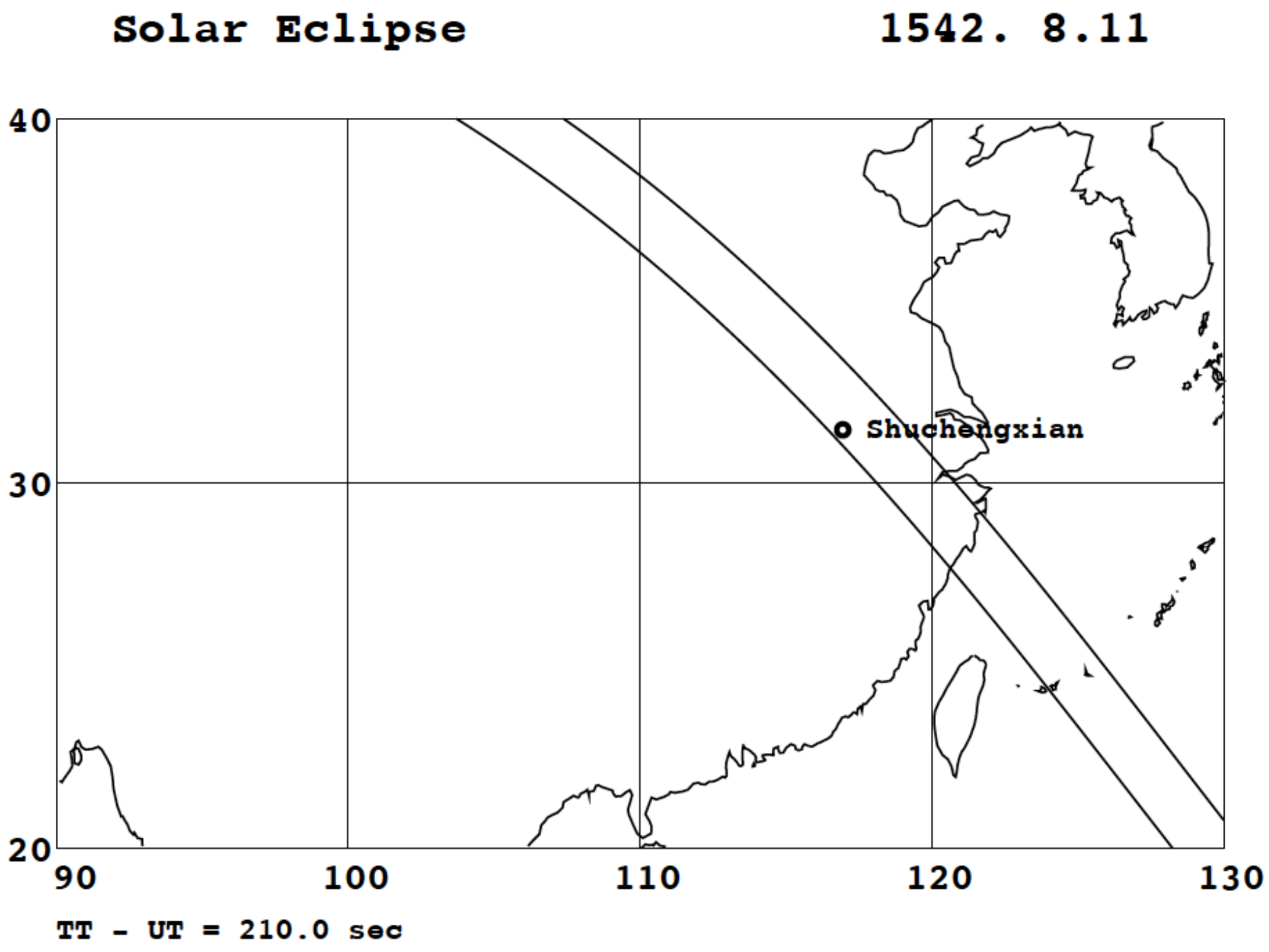

Figure 3: Our calculation of the totality path of the 1542 total solar eclipse in comparison with the location of Shūchéngxiàn, where we used M+21's ΔT value (ΔT = 210 s) in 1542.

Combining X09 and Google Earth Pro, we located Shūchéngxiàn at N31°28′, E116°57′ in the modern geographical coordinates, respectively. Using M+21's ΔT spline curve (ΔT = 210 s in 1542), our calculation located it in the totality path (Figure 3). In order to locate Shūchéngxiàn in the totality path, we need to locate the ΔT in 1542 in the range of -328 s ≤ ΔT ≤ 332 s. Our addition tightens the ΔT in 1542 as a new constraint against M+21.

Our result is contrasted with Han and Qiao (2003), who suggested a ΔT margin of 300 s < ΔT < 380 s in 1542 and whose result contradicted with M+21's ΔT spline curve (ΔT = 210 s in 1542). This is probably because what Han and Qiao (2003) used were mostly local treatises that were compiled during the Qīng period. Rather, our results set a firm upper limit in contrast with Han and Qiao (2003). If we follow BO88 and add Jiǔjiāngfǔzhì to our consideration, we can further narrow down the ΔT constraint. We locate the site at Jiǔjiāngfǔ (儀封縣; N34°48′, E114°56′) in the totality path too, if the ΔT stays in the margin of 148 s ≤ ΔT ≤ 332 s in 1542. Therefore, further investigation is needed.

**7. Total Solar Eclipse of 1575 May 10**

This is the only Míng eclipse that S97 analysed, and his results were not incorporated into subsequent studies (SMH16 and M+21). For this eclipse, we have located four local reports that explicitly mentioned totality and that was compiled during the Míng period. These records are read as follows:

R7-1: *Péngzéxiànzhì* (彭澤縣志 v. 7 f. 3b)
Our Transcription: 萬暦三年四月朔日，日食既，晝晦，鶏犬皆驚。
Our Translation: Wànlì 3rd year ... This year, 4th month, the first day; the Sun was totally eclipsed. The daylight became so dark that chickens and dogs were surprised.

R7-2: *Xiùshuǐxiànzhì* (秀水縣志, v. 10, f. 3a)
Our Transcription: 萬暦三年四月朔，日有食之既，晝晦，星見。
Our Translation: Wànlì 3rd year ... This year, 4th month, the first day; the Sun was totally eclipsed. Daylight became so dark that star(s) became visible.

R7-3: *Hǎiyánxiàn Tújīng* (海鹽縣圖經, v. 16, f. 7b)

Our Transcription: 萬曆三年… 是四月之朔，日有食之既。自午盡未，昏黑，不辨只尺，人情大駭。

Our Translation: Wànlì 3rd year ... This year, 4th month, the first day; the Sun was totally eclipsed. From wǔ to wèi, it became dark and gloomy; objects could not be distinguished even at a distance of a zhǐ or a chǐ, and the people were greatly alarmed.

R7-4: *Sōngjiāngfǔzhì* (松江府志 v. 47 f. 22b)

Our Transcription: 萬曆乙亥夏四月朔，日食。亭午食既，白晝如晦。

Our Translation: Wànlì Yǐhài year, summer 4th month, the first day; a solar eclipse took place at midday. It was a total eclipse. The daytime resembled night.

R7-5: *Míngshǐ* (明史 v. 20, p. 263)

Our Transcription: 萬曆三年…夏四月己巳朔，日有食之既。

Our Translation: Wànlì 3rd year, summer 4th month, on day jǐsì, the first day: the Sun was totally eclipsed.

Among these reports, R7-1 indicated abnormal animal reactions. R7-1, R7-2, R7-3, and R7-4 dated the preface or prints in Wànlì 10th year (≈ 1582), Wànlì 24th year (≈ 1596), Tiānqǐ 4th year (≈ 1624), and Chóngzhēn 4th year (≈ 1631), respectively. R7-1 was published only 7 years after this eclipse. The other records were published ≈ 21-56 years after the eclipse. For R7-1, it is possible to expect that either the author or one of their direct informants was an eyewitness to the eclipse. The others may rely their source in their hearsay, while we cannot exclude the possibility that their informants were direct eyewitnesses to this solar eclipse.

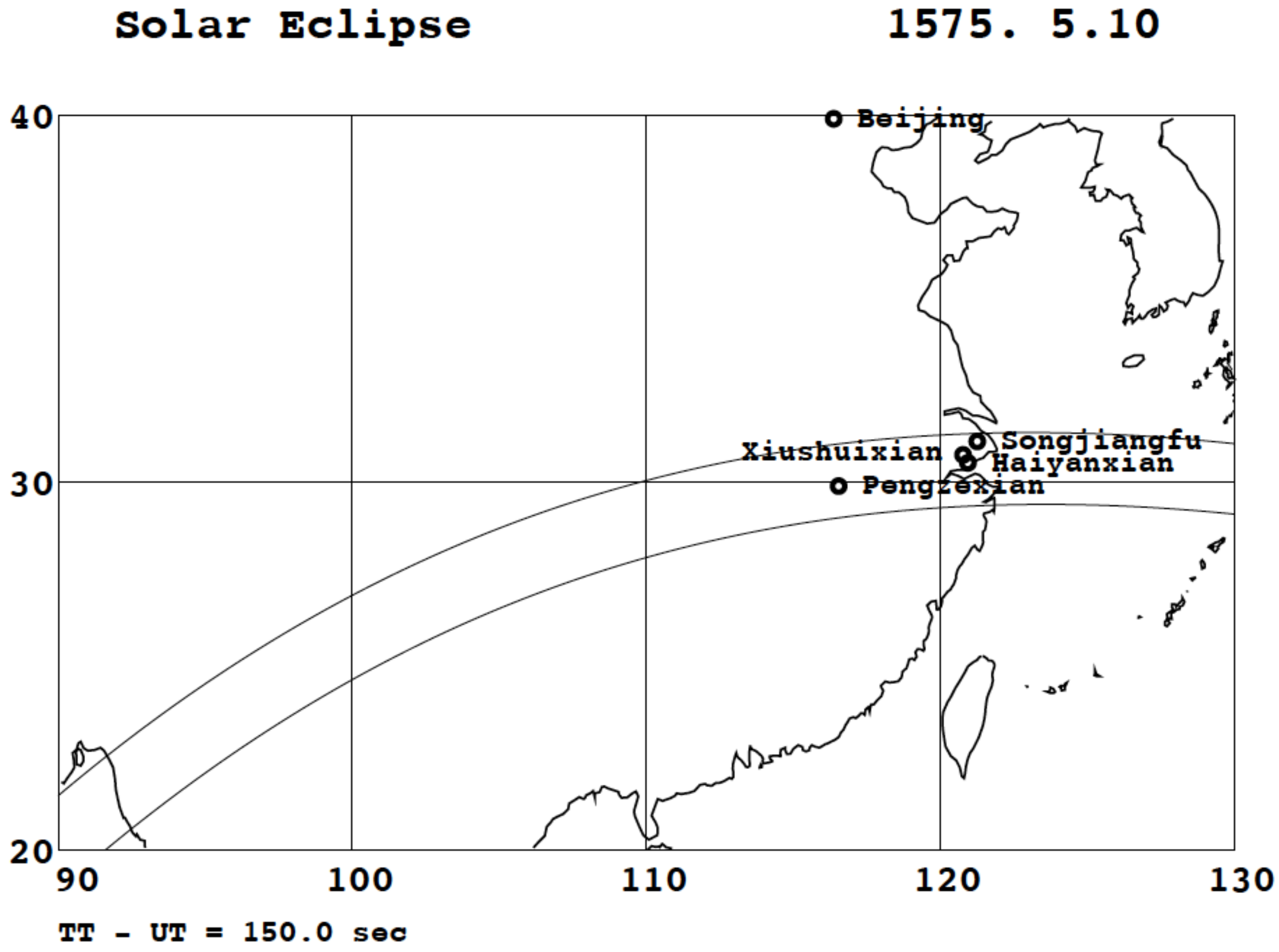

Figure 4: Our calculation of the totality path of the 1575 total solar eclipse in comparison with the location of Péngzéxiàn, Xiùshuǐxiàn, Hǎiyánxiàn, Sōngjiāngfǔ, and Běijīng, where we used M+21's ΔT value (ΔT = 150 s) in 1575.

Combining X09 and Google Earth Pro, we located Péngzéxiàn at N29°54′, E116°33′, Xiùshuǐxiàn at N30°45′, E120°47′, Hǎiyánxiàn at N30°32′, E120°57′, and Sōngjiāngfǔ at N31°07′, E121°16′, respectively. Using M+21's ΔT spline curve (ΔT = 150 s in 1575), our calculation located all these sites in the totality path (Figure 4). In order to locate these sites in the totality path, we need to locate the ΔT in 1575 in the range of −1762 s ≤ ΔT ≤ 1091 s. Our addition tightens the ΔT in 1575 in comparison with what S97 derived as −2150 s < ΔT < 1390 s and SMH16 and M+21 later missed for some reason. Around the China mainland, the total path ran almost in parallel to the equator. This made their relative location to the totality path less sensitive to the ΔT variation. Still, this is reasonably consistent with M+21's ΔT spline curve (ΔT = 150 s in 1575) and slightly tightens the ΔT variation around 1575.

Interestingly, R7-5 shows a case where *Míngshǐ*, the official history of the Míng Dynasty, mentioned a total obscuration in the Imperial Annals of Shénzōng (神宗本紀), in contrast with S97's explanation (S97, p. 260). As discussed in Dubs (1938, pp. 517-518) and Stephenson et al. (2018, pp. 427-428), eclipse reports in imperial annals did not come only from imperial observatories. This was also the case for *Míngshǐ*. During the reign of Shénzōng (神宗), the Míng Dynasty located the capital in Běijīng (北京: N39°54', E116°26'). As the totality path ran almost in parallel with the equator around the China mainland, it is unrealistic to locate Běijīng in the totality path. This case may serve as an additional proof for the caveat on usage of eclipse reports in the imperial annals and support discussions of Dubs (1938) and Stephenson et al. (2018).

**8. Revisiting the Contemporaneous ΔT Constraints**

Table 1 summarises our results from Sections 4 – 7. They need to be contextualised to the ΔT data from the known contemporaneous eclipse records. The previous studies up to M+21 offered 14 ΔT constraints in the 14th to 16th centuries from reports on the total/annular solar eclipses. Here, we critically revisit these existing ΔT constraints, following the exiting interpretations of eclipse records and revising the observational sites, using JPL DE 441 (Park et al., 2021) for the data homogeneity[3]. There results are summarised in Table 2.

For the 1310 ΔT constraint, M+21 used a Paris report of an annular solar eclipse and an Erfurt report of a partial solar eclipse on 31 January 1310. In order to locate Paris (N48°51' E002°21') within the totality path, we need to set a ΔT constraint of 498 s ≤ ΔT ≤ 2463 s. In order to locate Erfurt (N50°59' E011°02') out of the annularity path, we need to set a ΔT constraint of ΔT ≤ 1557 s. Their combination yields a ΔT constraint of 498 s ≤ ΔT ≤ 1557 s in 1310. This slightly improves M+21's existing ΔT constraint of 460 s < ΔT < 1555 s in 1310.

For the 1330 ΔT constraint, M+21 used a Zbraslav report of a partial eclipse on 16 July 1330, rejecting other records of Tüngedaer Schloß and Constantinople. The source report is acquired from *Chronicon Aulae Regiae*, by hand of local Cistercian abbots (Antonín, 2019). This monastery was located in the old town area of Zbraslav (N49°58.5' E014°23.5'). In order for this monastery to have witnessed the partial solar eclipse above the local horizon on 16 July 1330, two ΔT constraints are required for 1330: −5964 s ≤ ΔT ≤ 909 s and 1231 s ≤ ΔT ≤ 35691 s. The former is plausible, as the latter conflicts the contemporaneous ΔT constraints. This result confirms M+21's ΔT constraint of ΔT < 910 s and adds a non-critical lower ΔT limit.

[3] M+21 for example used NASA JPL DE 431 (Folkner et al., 2014) as their ephemeris dataset for their ΔT calculations. We need to reserve further discussions on the eclipse records in 1560 and 1567, as these cases require detailed analyses on the basis of the lunar-limb configurations.

For the 1378 ΔT constraint, M+21 used a Seville report for a total solar eclipse on 16 May 1379. We revised the coordinate of Seville (N37°23′, W006°00′), as M+21 located it in the in the southeastern vicinity of the actual old town area. In order to locate Seville in the totality path for this eclipse, we need to set a ΔT constraint of 107 s ≤ ΔT ≤ 1532 s. This slightly improves M+21's existing ΔT constraint of 105 s < ΔT < 1525 s in 1378.

For the 1386 ΔT constraint, M+21 used a Montpellier report for a total solar eclipse on 1 January 1386. In order to locate Montpellier (N43°36', E003°53') in the totality path for this eclipse, we need to set a ΔT constraint of −669 s ≤ ΔT ≤ 1074 s. This slightly improves M+21's existing ΔT constraint of −650 s < ΔT < 1075 s in 1386.

For the 1406 ΔT constraint, M+21 used a Bordeaux report and a Liege report for a total solar eclipse on 16 June 1406 and rejected a Hamburg report that S97 and Tanikawa et al. (2023, hereafter T+23) used. However, M+21 located Hamburg to N53°33', E007°38', near modern Wittmundhafen Air Base of Lower Saxony. This is some 160 km westward from Hamburg (N53°33', E009°59'). In order to locate Hamburg (N53°33', E009°59'), Liege (N50°38.5', E005°34.5'), and Bordeaux (N44°50', W000°34') in the totality path, we need to set ΔT constraints of −840 s ≤ ΔT ≤ 733 s, −612 s ≤ ΔT ≤ 876 s, and 396 s ≤ ΔT ≤ 1796 s, respectively. In combination, we need to set a ΔT constraint of 396 s ≤ ΔT ≤ 733 s in 1406 to satisfy these reports. This result substantially improves the existing ΔT constraints of 400 s < ΔT < 900 s (M+21) and 168 s < ΔT < 736 s (T+23), who used report combinations of Bordeaux-Liege and Magdeburg-Hamburg.

For the 1415 ΔT constraint, M+21 used a Wrocław report and a Kraków report for a total solar eclipse on 7 June 1415. However, Boyve (1855, p. 463) reported a total solar eclipse as seen from Neuchâtel[4] indicating an extraordinary darkness and abnormal bird behaviours. In order to locate

[4] The report reads: "Le 7 juin 1415 il y eut une éclipse de soleil si extraordinaire, 1415 qu'elle épouvanta les habitants de la terre. Sous l'horizon la nuit fut si obscure, que les oiseaux tombaient à terre" (Boyve, 1855, p. 463). We could translate this report as follows: "On 7 June 1415, there was a solar eclipse so extraordinary that it terrified the inhabitants on the Earth. Within the horizon, the night was so dark that the birds fell to the Earth".

Kraków (N50°3.5', E019°56'), Wrocław (N51°6.5', E017°02'), and Neuchâtel (N46°59.5', E006°55.5') in the totality path, we need to set ΔT constraints of 190 s ≤ ΔT ≤ 2024 s, −504 s ≤ ΔT ≤ 1330 s, and −1070 s ≤ ΔT ≤ 578 s, respectively. In combination, we need to set a ΔT constraint of 190 s ≤ ΔT ≤ 578 s in 1415 to satisfy these reports. This result substantially improves M+21 and T+23's existing ΔT constraints of 200 s < ΔT < 705 s and 45 s < ΔT < 675 s.

For the 1431 ΔT constraint, S97 used a Foligno report and a Perugia report for a total solar eclipse on 12 February 1431. SMH16 and M+21 commonly used this eclipse for one of their ΔT constraints. In order to locate Foligno (N42°57', E012°42') and Perugia (N43°06.5', E012°23.5') in the totality path, we need to set ΔT constraints of −18 s ≤ ΔT ≤ 843 s and −177 s ≤ ΔT ≤ 689 s, respectively. In combination, we need to set a ΔT constraint of −18 s ≤ ΔT ≤ 689 s in 1431 to satisfy these reports. This result slightly improves M+21 and T+23's existing ΔT constraints of −200 s < ΔT < 680 s and −16 s < ΔT < 679 s. M+21's value would be better explained, if SMH16 and M+21 rejected Foligno report for some reason.

For the 1433 ΔT constraint, M+21 used a Augsburg report and a Karlštejn report for a total solar eclipse on 17 June 1433. SMH16 and M+21 commonly use this eclipse for one of their ΔT constraints. In order to locate Augsburg (N48°22', E010°54') and Karlštejn (N49°56', E014°11') in the totality path, we need to set ΔT constraints of −995 s ≤ ΔT ≤ 575 s and 413 s ≤ ΔT ≤ 2034 s, respectively. In combination, we need to set a ΔT constraint of 413 s ≤ ΔT ≤ 575 s in 1433 to satisfy these reports. This result slightly improves M+21's existing ΔT constraint of 385 s < ΔT < 570 s, who located Karlštejn at Králův Dvůr (N49°57', E014°02'), some 11 km westward from the actual site. This is also contrasted with T+23's ΔT constraint of 300 s < ΔT < 445 s, who incorrectly assumed a partiality for an Aleppo report in contrast with the actual description.

For the 1478 ΔT constraint, M+21 used a Salamanca report for a total solar eclipse on 29 July 1478. Later, Martínez Usó and Marco Castillo (2023) added another report from Valencia. In order to locate Salamanca (N40°58', W005°40') and Valencia (N39°28.5', W000°22.5') in the totality path, we need to set ΔT constraints of −725 s ≤ ΔT ≤ 1106 s and −405 s ≤ ΔT ≤ 1201 s, respectively. In combination, we need to set a ΔT constraint of −405 s ≤ ΔT ≤ 1106 s in 1478 to satisfy these reports. This result slightly improves the existing ΔT constraints of −725 s < ΔT < 1095 s (M+21) and −450 s < ΔT < 1125 s (Martínez Usó and Marco Castillo, 2023).

For the 1485 ΔT constraint, M+21 used a Bourges report and a Fribourg report for a total solar eclipse on 16 March 1485. In order to locate Bourges (N47°05', E002°24') and Fribourg (N46°48', E007°10') in the totality path, we need to set ΔT constraints of −1867 s ≤ ΔT ≤ 497 s and −227 s ≤ ΔT ≤ 1955 s, respectively. In combination, we need to set a ΔT constraint of −227 s ≤ ΔT ≤ 497 s in 1485 to satisfy these reports. This result slightly improves M+21's existing ΔT constraints of −215 s < ΔT < 520 s.

For the 1539 ΔT constraint, M+21 used a Seville report for a total solar eclipse on 18 April 1539. In order to locate Seville (N37°23′, W006°00′) in the totality path, we need to set a ΔT constraint of −66 s ≤ ΔT ≤ 2316 s. Note that M+21 used a different ephemeris dataset and located Seville in the southeastern vicinity of the actual old town area. During this eclipse, the totality path ran almost parallel to the equator in the European sector. This made their relative location to the totality path less sensitive to the ΔT variation. This result slightly improves M+21's existing ΔT constraints of −10 s < ΔT < 2345 s.

For the 1598 ΔT constraint, M+21 used an Edinburgh report and a St. Andrews report for a total solar eclipse on 7 March 1598. In order to locate Edinburgh (N55°57′, W003°12′) and St. Andrews (N56°20.5', W002°47.5') in the totality path for this eclipse, we need to set ΔT constraints of −385 s ≤ ΔT ≤ 203 s and −343 s ≤ ΔT ≤ 246 s. In combination, we need to set a ΔT constraint of −343 s ≤ ΔT ≤ 203 s in 1598 to satisfy these reports. This slightly improves M+21's existing ΔT constraint of −360 s < ΔT < 245 s in 1598. In contrast with M+21, we used the Edinburgh report for the upper ΔT limit and the St. Andrews report for the lower ΔT limit.

## 9. Summary and Discussions

Figure 5 contextualises our results into the contemporaneous ΔT data (Table 1) into M+21's existing ΔT constraints and ΔT spline curve, revising one ΔT constraint (1361) and adding three new ΔT constraints (1514, 1542, and 1575). Our revision of the 1361 eclipse reports requires us to either revise M+21's ΔT spline curve upward (from ΔT = 557 s to 602 s ≤ ΔT ≤ 1864 s) or change the interpretation for the local eclipse visibility from totality (Morrison et al., 2020; M+21) to a deep partiality (S97; SMH16). We incline towards the latter scenario and suggest a revised ΔT constraint of −408 s ≤ ΔT ≤ 601 s in 1361.

Table 1: Critical ΔT limits and their sources discussed in this study, showing the date of the eclipse, the observational sites, their geographical coordinates, M+21's ΔT value, our ΔT constraints, and references.

| Date | Site | Latitude | Longitude | M+21's ΔT (s) | Our ΔT constraints (s) | Sources |
|---|---|---|---|---|---|---|
| 1361 May 5 | Sōngjiāngfǔ | N31°07′ | E121°16′ | 557 | −408 s ≤ ΔT ≤ 601 s | Section 4 |
| 1361 May 5 | Sōngjiāngfǔ | N31°07′ | E121°16′ | 557 | 1865 s ≤ ΔT ≤ 18532 s | Section 4 |
| 1514 Aug 20 | Jiǔjiāngfǔ | N29°42′ | E115°59′ | 264 | 277 s ≤ ΔT ≤ 1627 s | Section 5 |
| 1514 Aug 20 | Dōngxiāngxiàn | N28°14′ | E116°36′ | 264 | −347 s ≤ ΔT ≤ 890 s | Section 5 |
| 1542 Aug 11 | Shūchéngxià | N31°28′ | E116°57′ | 210 | −328 s ≤ ΔT ≤ 332 s | Section 6 |
| 1575 May 10 | Sōngjiāngfǔ | N31°07′ | E121°16′ | 150 | −1762 s ≤ ΔT ≤ 1091 s | Section 7 |

Table 2: All the ΔT constraints discussed in this study, showing the eclipse date, our site choice, and the ΔT constraints of ours, M+21, and other previous studies. We remark S97 (his Appendix B) with *, Tanikawa et al. (2023) with *, Martínez Usó and Marco Castillo (2023) with **, and Han and Qiao (2003) with ***. When we have overlaps in "other previous studies", we showed the latest values among them. We omitted what S97 doubted the reliability of the source report. The entries with # are reserved for future studies.

| Eclipse Date | Our Site Choice | Ours | M+21 | Other Previous Studies |
|---|---|---|---|---|
| 1310 Jan 31 | Paris<br>Erfurt | 498 s ≤ ΔT ≤ 1557 s | 460 s < ΔT < 1555 s | |
| 1330 Jul 15 | Zbslav | −5964 s ≤ ΔT ≤ 909 s | ΔT < 910 s | ΔT < 890 s* |
| 1330 Jul 15 | Zbslav | 1231 s ≤ ΔT ≤ 35691 s | | 1210 s < ΔT* |
| 1361 May 5 | Sōngjiāngfǔ | −408 s ≤ ΔT ≤ 601 s | 500 s < ΔT < 1760 s | ΔT < 500 s* |
| 1361 May 5 | Sōngjiāngfǔ | 1865 s ≤ ΔT ≤ 18532 s | | 1760s < ΔT* |
| 1379 May 16 | Seville | 107 s ≤ ΔT ≤ 1532 s | 105 s < ΔT < 1525 s | |
| 1386 Jan 1 | Montpellier | −669 s ≤ ΔT ≤ 1074 s | −650 s < ΔT < 1075 s | |
| 1406 Jun 16 | Bordeaux<br>Hamburg | 396 s ≤ ΔT ≤ 733 s | 400 s < ΔT < 900 s | 170 s < ΔT < 740 s*<br>168 s < ΔT < 736 s** |
| 1415 Jun 7 | Neuchâtel<br>Kraków | 190 s ≤ ΔT ≤ 578 s | 200 s < ΔT < 705 s | −650 s < ΔT < 670 s*<br>45 s < ΔT < 675 s** |
| 1431 Feb 12 | Foligno<br>Perugia | −18 s ≤ ΔT ≤ 689 s | −200 s < ΔT < 680 s | 10 s < ΔT < 700 s*<br>−16 s < ΔT < 679 s** |
| 1433 Jun 17 | Augsburg<br>Karlštejn | 413 s ≤ ΔT ≤ 575 s | 385 s < ΔT < 570 s | 300 s < ΔT < 445 s** |
| 1478 Jul 29 | Salamanca<br>Valencia | −405 s ≤ ΔT ≤ 1106 s | −725 s < ΔT < 1095 s | −450 s < ΔT < 1125 s*** |
| 1485 Mar 16 | Bourges<br>Fribourg | −227 s ≤ ΔT ≤ 497 s | −215 s < ΔT < 520 s | −5500 s < ΔT < 780 s* |
| 1514 Aug 20 | Jiǔjiāngfǔ<br>Dōngxiāngxiàn | 277 s ≤ ΔT ≤ 890 s | | |
| 1539 Apr 18 | Seville | −66 s ≤ ΔT ≤ 2316 s | −10 s < ΔT < 2345 s | |
| 1542 Aug 11 | Shūchéngxià | −328 s ≤ ΔT ≤ 332 s | | 300 s < ΔT < 380 s**** |
| 1560 | Coimbra | # | −480 s < ΔT < 210 s | −475 s < ΔT < 205 s* |
| 1567 | Rome | # | 145 s < ΔT < 165 s | 145 s < ΔT < 165 s* |
| 1575 May 10 | Sōngjiāngfǔ | −1762 s ≤ ΔT ≤ 1091 s | | −2150 s < ΔT < 1390 s* |
| 1598 Mar 7 | St Andrews<br>Edinburgh | −343 s ≤ ΔT ≤ 203 s | −360 s < ΔT < 245 s | |

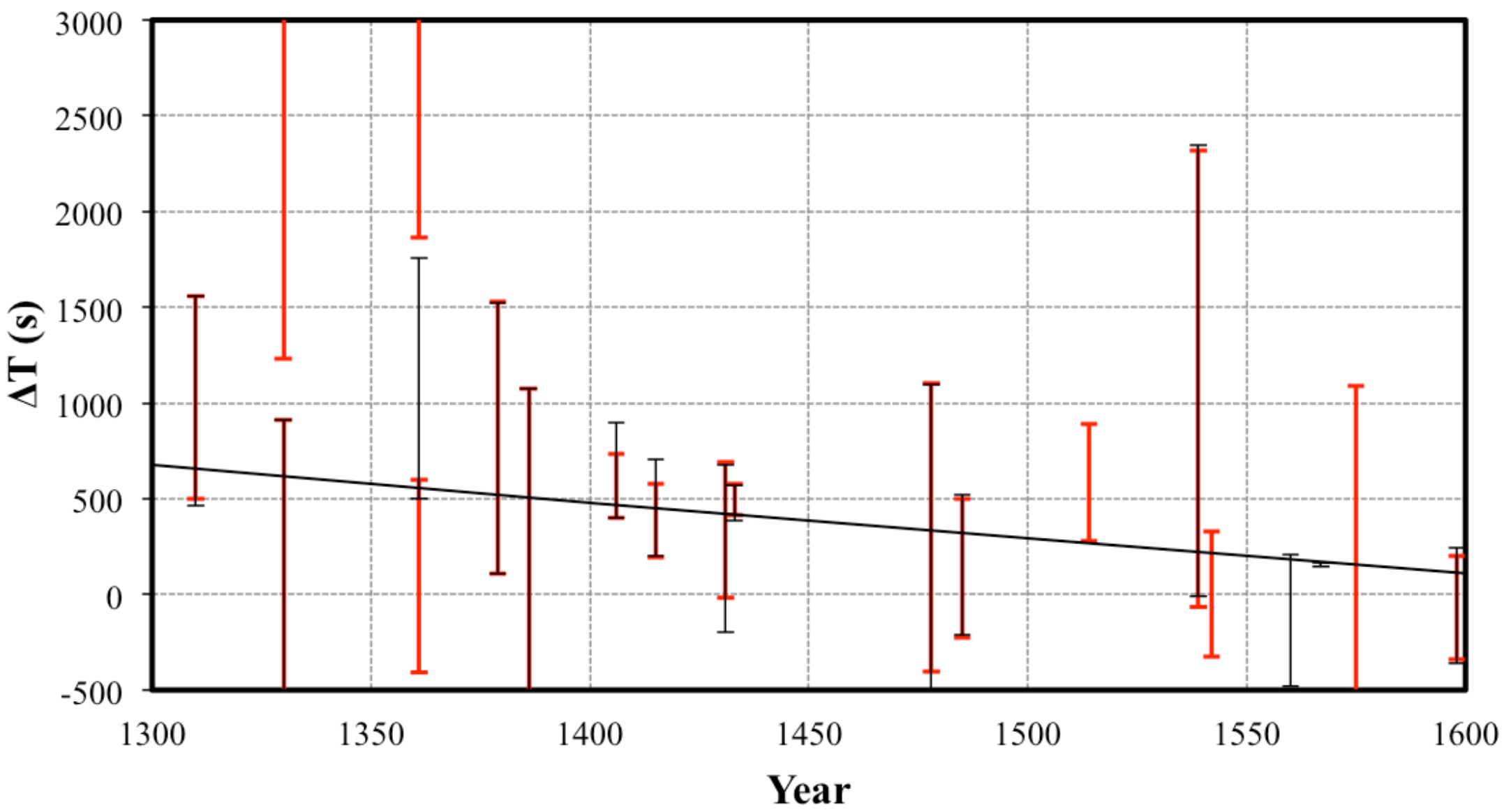

Figure 5: Comparison of the ΔT constraints that we derived from the Chinese local treatises in this study (red bars) and what M+21 showed in their supplementary table (black bars). The solid black curve shows M+21's ΔT spline curve. We show two ΔT constraints for the 1361 eclipse, as we interpret this eclipse as a deep partial solar eclipse rather than a total solar eclipse.

For the early 16th century, our study added two new ΔT constraints: 277 s ≤ ΔT ≤ 890 s in 1514 and −328 s ≤ ΔT ≤ 332 s in 1542. These values compare with the only ΔT constraint of −66 s ≤ ΔT ≤ 2316 s in 1539 that we revised from M+21 (Table 2). As such, our new ΔT constraints dramatically tighten the ΔT variations in the early 16th century and even require revising the ΔT around 1514 upward. For the late 16th century, our new ΔT constraint in 1575 (−1762 s ≤ ΔT ≤ 1091 s) only slightly tightens the ΔT variability. Still, this is a new ΔT constraint and independently supports M+21's existing ΔT constraints and ΔT spline curve around this period.

Overall, our ΔT constraints generally tighten the ΔT variations more than what M+21 fit for their ΔT spline curve, requiring revisions to the 1361 eclipse details and upward modifications for the ΔT reconstructions around 1542. This may indicate that the ΔT decrease between 1514 and 1567 was slightly steeper than previously considered. Further analyses of past eclipse records are required to tighten or revise past ΔT variations. Indeed, there is much to learn from the past.

Some of the critical ΔT constraints may allow us to figure out the variation of the length of day (LOD), if we assume the ΔT does not increase in annual to decadal timescales. Section 4 set the upper ΔT limit of 1361 as ΔT ≤ 601 s. Subsequently, the lower ΔT limits are set as 396 s ≤ ΔT in 1406 and 413 s ≤ ΔT in 1433. In combination, these data allow us to tentatively set the actual ΔT values in 1361 as 413 s ≤ ΔT ≤ 601 s.

Section 5 set the lower ΔT limit of 1514 as 277 s ≤ ΔT. Beforehand, the upper ΔT limit is set as ΔT ≤ 497 s in 1485. This value should not have increased by 1514. In combination, these data allow us to tentatively set the actual ΔT values in 1514 as 277 s ≤ ΔT ≤ 497 s. Section 6 set the upper ΔT limit of 1542 as ΔT ≤ 332 s. Beforehand, the lower ΔT limit is set as 277 s ≤ ΔT in 1514. In combination, these data allow us to tentatively set the actual ΔT values in 1542 as 277 s ≤ ΔT ≤ 332 s. These constraints are compared with the ΔT constraint of 145 s ≤ ΔT ≤ 165 s in 1567 (Stephenson et al., 1997; M+21).

This study allows us to detect short-term ΔT fluctuations that SMH16 and M+21's ΔT spline curves have missed so far. It is desired to add further ΔT constraints and tighten the ΔT variations in the past so that we can compare the eclipse-based LOD measurements with the sophisticated LOD modelings for the geodynamo activity (*e.g.*, Kiani-Shahvandi et al., 2024; Suttie et al., 2025). In the same time, we need to be cautious on the eclipse-based LOD measurements, as those before the timed occultation records are feasible only with a series of tight ΔT constraints. It is extremely important to examine multiple eclipse reports for each TSE and tighten the ΔT constraints as much as possible for each case. Accumulating their case studies will eventually offers us hints on future comparison with the model led LOD variations.

**Data Availability**

The source records are presented in the Appendix. Their copies were obtained from the National Central Library of Taipei. We used the ephemeris data from NASA JPL DE 441 (Park et al., 2021).

**Acknowledgments**

We wish to thank the National Central Library of Taipei for allowing us to access the source records and the Kanseki Database of Kyoto University for accessing their bibliographical details. This research was conducted under the financial support of JSPS Grant-in-Aids JP25K17436 and JP25H00635, the ISEE director's leadership fund for FYs 2021–2026, the Young Leader Cultivation (YLC) programme of Nagoya University, Tokai Pathways to Global Excellence (Nagoya University) of the Strategic Professional Development Program for Young Researchers (MEXT), the young researcher units for the advancement of new and undeveloped fields in Nagoya University Program for Research Enhancement, and the NIHU Multidisciplinary Collaborative Research Projects NINJAL unit "Rediscovery of Citizen Science Culture in the Regions and Today". Eclipse calculations were partly carried out on the Multi-wavelength Data Analysis System operated by the Astronomy Data Center, National Astronomical Observatory of Japan.

**References**

Antonín, R. 2019, *Chronica Aulae regiae*—an Unsuccessful Attempt to Establish an Official Memory of the Last Přemyslids and the Zbraslav Monastery, *The Medieval Chronicle*, **12**, 1-23

Ba, Z. 2015, Shìshù Dàmíng Yītŏngzhì De Kānbĕn Jí Qí Lìshĭ Gòngxiàn, *Zhōngguó Dìfāngzhì*, **1**, 28-36 [in Chinese]

Beijing Observatory, 1988, *Zhōngguó Gǔdài Tiānxiàng Jìlù Zǒngjí*, Nánjīng, Jiāngsū Kēxué Jìshù Chūbǎnshè [in Chinese] (BO88)

Boyve, J. 1854, *Annales historiques du Comté de Neuchâtel et Valangin depuis Jules-César jusqu'en 1722*, Berne, E. Mathey

Brooks, T., 1988, *Geographical Sources of Ming-Qing Histories*, Ann Arbor, The University of Michigan Press

Dubs, H. H. 1938, Solar Eclipses during the Former Han Period, *Osiris*, **5**, 499-522

Espenak, F., Meeus, J. 2009, *Five Millennium Catalog of Solar Eclipses: –1999 to +3000 (2000 BCE to 3000 CE)—Revised (NASA/TP–2009–214174)*, Greenbelt, NASA.

Fiala, A. D., Dunham, D. W., Sofia, S. 1994, Variation of the Solar Diameter from Solar Eclipse Observations, 1715 – 1991, *Solar Physics*, **152**, 97-104. DOI: 10.1007/BF01473190

Folkner, W. M., Williams, J. G., Boggs, D. H., Park, R. S., Kuchynka, P.: 2014, *IPN Progress Report*, **2014-02-15**, 42-196.

Han, Y.-B., Qiao, Q.-Y. 2003, A Check on the Variations of Earth's Rotation with an Ancient Solar Eclipse, *Chinese Journal of Astronomy & Astrophysics*, **3**, 569-575. DOI: 10.1088/1009-9271/3/6/569

Hayakawa, H., Lockwood, M., Owens, M. J., Sôma, M., Besser, B. P., van Driel, L. 2021, Graphical Evidence for the Solar Coronal Structure during the Maunder Minimum: Comparative Study of the Total Eclipse Drawings in 1706 and 1715, *Journal of Space Weather and Space Climate*, **11**, 1. DOI: 10.1051/swsc/2020035

Hayakawa, H., Murata, K., Sôma, M. 2022, The Variable Earth's Rotation in the 4th–7th Centuries: New ΔT Constraints from Byzantine Eclipse Records, *Publications of the Astronomical Society of the Pacific*, **134**, 094401. DOI: 10.1088/1538-3873/ac6b56

Hayakawa, H., Murata, K., Owens, M. J., Lockwood, M. 2024, Analyses for graphical records for a total solar eclipse in May 1230: a possible reference for the “Medieval Grand Maximum”, *Monthly Notices of the Royal Astronomical Society*, **530**, 3150-3159. DOI: 10.1093/mnras/stad3874

Hayakawa, H., Owens, M. J., Meng, J., Sôma, M., Lockwood, M. 2025, Analyses of the Ancient Chinese Report on the Total Solar Eclipse in 709 BCE: Implications for the Contemporaneous Earth’s Rotation Speed and Solar Cycles, *The Astrophysical Journal Letters*, **995**, L1. DOI: 10.3847/2041-8213/ae0461

Hucker, C. O. 1958, Governmental Organization of the Ming Dynasty, *Harvard Journal of Asiatic Studies*, **21**,1-66

Kiani Shahvandi, M., Noir, J., Mishra, S., Soja, B. 2024, Length of Day Variations Explained in a Bayesian Framework, *Geophysical Research Letters*, 51, 2024GL111148. DOI: 10.1029/2024GL111148

Lambeck, K., Rouby, H., Purcell, A., Sun, Y., Sambridge, M. 2014, Sea Level and Global Ice Volumes from the Last Glacial Maximum to the Holocene, *Proceedings of the National Academy of Sciences*, **111**, 15296-15303. DOI: 10.1073/pnas.1411762111

Littmann, M., Espenak, F. 2017, *Totality: the great American eclipses of 2017 and 2024*, Oxford: Oxford University Press

Loucif, M. L., Koutchmy, S. 1989, Solar cycle variations of coronal structures, *Astronomy and Astrophysics Supplement Series*, **77**, 45-66

Martínez Usó, M. J., Marco Castillo, F. J. 2023, The Total Eclipse of the Sun of July 29, AD1478, in Contemporary Spanish Documents, *Journal for the History of Astronomy*, **54**, 153-170. DOI: 10.1177/00218286231167157

Mitrovica, J. X., Hay, C. C., Morrow, E., Kopp, R. E., Dumberry, M., Stanley, S. 2015, Reconciling Past Changes in Earth's Rotation with 20th Century Global Sea-Level Rise: Resolving Munk's Enigma, *Science Advances*, **1**, e1500679. DOI: 10.1126/sciadv.1500679

Morrison, L. V., Stephenson, F. R., Hohenkerk, C. Y. 2020 Historical Changes in UT/LOD from Eclipses, *Journées*, **2017**, 167-172. See https://web.ua.es/journees2017/proceedings/PROCEEDINGS-JOURNEES.pdf

Morrison, L. V., Stephenson, F. R., Hohenkerk, C. Y., Zawilski, M. 2021, Addendum 2020 to `Measurement of the Earth's rotation: 720 BC to AD 2015', *Proceedings of the Royal Society A*, **477**, 20200776. DOI: 10.1098/rspa.2020.0776 (M+21)

Orchiston, W., Green, D. A., Strom, R. 2015, *New Insights From Recent Studies in Historical Astronomy: Following in the Footsteps of F. Richard Stephenson: A Meeting to Honor F. Richard Stephenson on His 70th Birthday*, Cham: Springer International Publishing

Park, R. S., Folkner, W. M., Williams, J. G., Boggs, D. H. 2021, The JPL Planetary and Lunar Ephemerides DE440 and DE441, *The Astronomical Journal*, **161**, 105. DOI: 10.3847/1538-3881/abd414 (P+21)

Pasachoff, J. M. 2017, Heliophysics at Total Solar Eclipses, *Nature Astronomy*, **1**, 0190. DOI: 10.1038/s41550-017-0190

Rekier, J., Chao, B. F., Chen, J., Dehant, V., Rosat, S., Zhu, P. 2022, Earth's Rotation: Observations and Relation to Deep Interior,

Riley, P., Lionello, R., Linker, J. A., et al. 2015, Inferring the Structure of the Solar Corona and Inner Heliosphere During the Maunder Minimum Using Global Thermodynamic Magnetohydrodynamic Simulations, *The Astrophysical Journal*, **802**, 105. DOI: 10.1088/0004-637X/802/2/105

Rozelot, J. P., Damiani, C. 2012, Rights and Wrongs of the Temporal Solar Radius Variability, *The European Physical Journal H*, **37**, 709-743. DOI: 10.1140/epjh/e2012-20030-4

Stephenson, F. R. 1997, *Historical Eclipses and Earth's Rotation*, Cambridge, Cambridge University Press (S97)

Stephenson, F. R., Morrison, L. V., Hohenkerk, C. Y. 2016, Measurement of the Earth's rotation: 720 BC to AD 2015, *Proceedings of the Royal Society A*, **472**, 20160404. DOI: 10.1098/rspa.2016.0404 (SMH16)

Stephenson, F. R., Morrison, L. V., Hohenkerk, C. Y. 2018, The Provenance of Early Chinese Records of Large Solar Eclipses and the Determination of the Earth's Rotation, *Journal for the History of Astronomy*, **49**, 425-471. DOI: 10.1177/0021828618789850

Suttie, N., Nilsson, A., Gillet, N., Dumberry, M. 2025, Large-Scale Palaeoflow at the Top of Earth's Core, *Earth and Planetary Science Letters*, **652**, 119185. DOI: 10.1016/j.epsl.2024.119185

Sôma, M., Tanikawa, K.: 2015, Determination of ΔT and Lunar Tidal Acceleration from Ancient Eclipses and Occultations, in: W. Orchiston et al. (eds.), *New Insights From Recent Studies in Historical Astronomy: Following in the Footsteps of F. Richard Stephenson,* Springer, Cham, pp. 11-23. DOI: 10.1007/978-3-319-07614-0_2

Tanikawa, K., Sôma, M., Simmonds, O., Iwahashi, K. 2023, Solar Eclipses Observed Worldwide in the First Half of the 15th Century and ΔT Determined from Multiply Observed Eclipses, *Publications of the Astronomical Society of Japan*, **75**, 1013-1029. DOI: 10.1093/pasj/psad036

Xue, G. 2009, *Zhōngguó Gǔjīn Demíng Duìzhàobiǎo*, Shànghǎi, Shànghǎi Císhū Chūbǎnshè [in Chinese]

Zhu, P., Triana, S. A., Rekier, J., Trinh, A., Dehant, V. 2021, Quantification of Corrections for the Main Lunisolar Nutation Components and Analysis of the Free Core Nutation from VLBI-Observed Nutation Residuals, *Journal of Geodesy*, **95**, 57. DOI: 10.1007/s00190-021-01513-9

**Appendix: Source Documents**

They are shown in their original languages with a reference ID in the National Central Library of Taipei (NCL). Their copies are also found elsewhere. See Kanseki Database[5] to locate their copies in Japanese archives.

*Dōngxiāngxiànzhì*: 東鄉縣志, NCL 672.49/405.1 676

*Hǎiyánxiàn Tújīng*: 海鹽縣圖經, NCL 672.39/123.1 6763 2014 v. 4

*Jiǔjiāngfǔzhì*: 九江府志, NCL 672.49/400.1 678

*Míngshǐ*: 明史, 中華書局, 1975

*Míngshīzōng*: 明詩綜, 卷十二・陶宗儀 NCL R 082.15 7596 75 v.491

*Náncūn Chuògēnglù*: 南村輟耕錄, NCL R 082.1 8367 72 v.1040

*Péngzéxiànzhì*: 彭澤縣志, NCL 210.2 03517

*Shūchéngxiànzhì*: 舒城縣志, NCL 210.2 03458

*Sōngjiāngfǔzhì*: 松江府志, NCL 672.19/204.1 6573-2 v. 12

*Xiùshuǐxiànzhì*: 秀水縣志, NCL 672.35/119 644 v. 2

5 http://kanji.zinbun.kyoto-u.ac.jp/kanseki